# Winning Over Future Scientists

Julie Fry
Department of Physics and Astronomy and Department of Chemistry
University of Rochester, Rochester, NY 14627
UR- 1582- September 1999
(A revised version of this article will be submitted
to Journal of Women and Minorities in Science)

Abstract

A review of existing pre-college science programs for young women in high school is presented, with emphasis on the University of Rochester Pre-College Experience in Physics summer program for 9$^{th}$ and 10$^{th}$ grade high school women (PREP-CMS). A new model for such programs is proposed.

In a world that is increasingly dependent on science and technology, the United States is not capitalizing on all of its national resources, certainly not those of the mind. Scientific research has been historically a male pursuit, with a few "female geniuses" venturing into the intellectual domain of their fathers and brothers to contribute to the scientific community. Schools are working to change‹ this. Students read about more of their notable foremothers like Marie Curie, Elizabeth Blackwell, Jane Goodall, Chien-shiung Wu, and Maria Mitchell, and they are viable role models, not the exception to the rule. In their textbooks they see pictures holding bubbling beakers. As a society, we have decided that women can do science, and a slow process of change has begun. Women in college physics and chemistry courses have more and more female classmates, but in 1993, women received only 35% of the physical science bachelor's degrees, and about 16% of engineering bachelor's degrees [1]. For comparison, women were 73% of psychology bachelor's degree recipients Increasing numbers of women in scientific work and academia expand the possibilities for breakthrough and broaden horizons in teaching and research methods. It is also increasingly being acknowledged as a business advantage to strive for diversity in employment practices. This changing perception in business is naturally felt by scientific/technological professions as well. Acknowledging that this promotion of women into science is a desirable progression, I will address in this paper the issues surrounding the pedagogical and extracurricular efforts to attract high school girls to science, and develop a unique proposal to accelerate the influx of young women into the natural sciences.

## The Problem

The first question to address en route to a better understanding of what can be done to increase the numbers of women studying science in college is the question of the possible causal factors of the underrepresentation. Why are women underrepresented in university natural science curricula? There are diverse answers to this question, many of which depend on geography, family structure, and economic and social status. These represent the situation into which girls are born, and are beyond the scope of this paper. The remaining major factor is the focus of this discussion: education outside of the home.

      Middle school and high school are the primary social venues for girls and boys, where they learn many of the lifeskills that will serve them in whatever they pursue later in life.  As their teachers strive to provide gender-equitable classrooms, to varying degrees of success, the students not only learn about History and English and Science, but are also subjected to what has been referred to as the "hidden curriculum".  The most persistent lesson in this hidden curriculum is the one pointed out in Myra and David Sadker's studies on sexism in the classroom:  boys were consistently asked more complex questions and praised for their academic ability, while girls were usually commended for their good behavior and docility. (2,3)  It is through subtle lessons like these that a group of students develop gender differentiations that can eventually contribute to the discouragement of women in science and math.

      In recent years, there has been a groundswell of research on girls in schools, most notably the AAUW Report, "How Schools Shortchange Girls".  This extensive national survey showed that for girls, the passage into adolescence was marked not only by changes in their bodies, but also by a loss of confidence in self and abilities, especially in math and science.  This "confidence gap" immediately precedes the drop in achievement that is observed in standardized test scores in high school.  The confidence gap is a somewhat fuzzy concept, so it is important to explain how this drop in confidence is measured.  The decline in self-esteem and self-confidence as girls and boys move from childhood into adolescence is measured by large-scale empirical studies, public-opinion polls, but most convincingly by in-depth longitudinal clinical studies that follow individual groups of girls and boys through school.  One comprehensive survey study commissioned by the AAUW in 1990 found that on average, 69% of boys and 60% of girls in elementary school agreed with the statement, "I am happy the way I am."  The same survey administered to high school students at the same time showed 46% of boys and only 29% of girls "happy the way I am". (4)  Thus it is this confidence gap, and not an inherent ability gap, that may explain the disproportionate ratio of men to women interested in the physical sciences.  The subsequent drop in standardized test scores is a meaningful and frightening trend.  It cannot be brushed aside based on presumed gender-bias of standardized tests, because if there were bias, it would naturally cause a relatively consistent gap in girls' and boys' scores throughout school.  The trend that is observed is a widening of that gap between the elementary school standardized tests and the high school testing, in spite of the fact that all tests are designed and administered by the same Educational Testing Service. (5)  High school students often perceive the SAT score as a very significant indicator of their abilities, and this gap in achievement on the SAT can further discourage girls from pursuing fields in college that they perceive as most challenging.

      According to Peggy Orenstein, the author of a 1994 study on SchoolGirls in two California middle schools, "We live in a culture that is ambivalent toward female achievement, proficiency, independence, and right to a full and equal life." (6)  The consequences of this cultural prejudice are successive generations of women who, although they face fewer tangible career barriers than their mothers and grandmothers, still enter adulthood with a weaker sense of self than their male

classmates. This lack of confidence makes women less willing to pursue non-traditional careers and prevents anything but a gradual increase in numbers of women scientists.

Educators have been concerning themselves with this issue for several years. The first international Girls and Science and Technology conference met in 1981 (7), and since then there has been a continuing dialogue about understanding and effectively addressing the barriers, whether subtle or blatant, to women in science education. Researchers have studied the historical construction of science, observed classrooms, and analyzed curricula and teaching methods in search of sources of gender inequity. (1,2,3,6,8,12,13,14)

The most well-known and exhaustive research on gender inequity in the classroom was commissioned by the American Association of University Women in 1990. The resulting publication, "How Schools Shortchange Girls -- The AAUW Report" received much public attention and in many cases was met with genuine disbelief. For many, it amounted to a call to action, and inspired further research and serious scrutiny of accepted classroom interactions and teaching methods. The Report was compiled from more than 1300 studies at schools across the nation, documenting a problem of national proportions. The studies were rich with evidence that "girls are not receiving the same quality, or even quantity, of education as their brothers." (5) Most importantly, it follows up on the disturbing evidence of a dismal situation in the nation's schools with recommendations to rectify the situation and be sure that both boys and girls are well equipped to maximize their potential when they graduate from high school. The conclusion of the AAUW Educational Foundation President, Alice Ann Leidel, concurs with the pragmatic reasoning of the discussion here: "To remain competitive in the global economy, we need to build the skills of all our children. If we shortchange girls, we shortchange America." (8)

The National Science Foundation (NSF) echoes an expanding group of business professionals in articulating this very pragmatic reason to encourage more women to take science courses: the demand by the job market for an increasingly diverse labor force. Businesses are increasingly noting the positive effect diversity in their employees has on the bottom line. The US Bureau of Labor Statistics projects that as overall rates of entry into the labor force are decreasing between 1994 to 2005, the women's labor force will grow twice as quickly as the men's labor force. Combining this projection with the prediction that by 2005, the high-tech jobs the new labor force will be preparing for will require higher level skills in science, engineering, and mathematics than ever before is a convincing practical argument that we need to actively work to change the low participation of women in these fields. (9)

NSF outlines its version of the most important issues relating to the national problem in the following way:—

-The disproportionately high numbers of girls who lose interest in science during elementary and middle school;

-The low numbers of women who enroll in advanced high school science and math courses to prepare for college;

-The disproportionately low numbers of women entering undergraduate studies in SEM, particularly in physical sciences, computer sciences,

and engineering;

-The low number of women completing SEM graduate degrees; and

-The slow rate of women's advancement to senior ranks and leadership positions in academic, business, and government careers. (9)

Through numerous studies, many of which the AAUW cited in the Report, it has been firmly established that young women are not receiving the same teacher attention, are not seeing viable future "selves" in their textbooks and media images of scientists, are not having as much expected of them in math and science, and are less frequently encouraged to take higher-level math and science courses. This suggests a necessity for institutional change. (5,9) The AAUW Report concludes with a lengthy list of recommendations, 10% of which are focused specifically on the areas of science and math, where the studies found the most disparity. The specific recommendations in this area, echoed in the popular media and many other academic publications, are as follows:

> Girls must be educated and encouraged to understand that mathematics and the sciences are important and relevant to their lives. Girls must be actively supported in pursuing education and employment in these areas.
>
> 17. The federal government must fund and encourage research on the effect on girls and boys of new curricula in the sciences and mathematics. Research is needed particularly in science areas where boys appear to be improving their performance while girls are not.
>
> 18. Educational institutions, professional organizations, and the business community must work together to dispel myths about math and science as "inappropriate" fields for women.
>
> 19. Local schools and communities must encourage and support girls studying science and mathematics by showcasing women role models in scientific and technological fields, disseminating career information, and offering "hands-on" experiences and work groups in science and math classes.
>
> 20. Local schools should seek strong links with youth-serving organizations that have developed successful out-of-school programs for girls in mathematics and science and with those girls' schools that have developed effective programs in these areas. (5)

School districts will continue to discuss how they will ensure that girls are being encouraged and supported in the science and math classrooms. Academics will conduct further studies to understand more fully why girls are falling behind in certain subjects, and promote awareness of these problems.  The remaining avenue for the encouragement and education of young women in science will be addressed by this paper: the out-of-school programs.

**Solutions**

Fortunately, wealthy organizations like the NSF don't just outline the problems to tell women how they're slighted, they also demand change and offer money to researchers and departments to affect that change.  The groundswell of research on gender differences in education, and particularly the shortfall in science education, has been answered by the development of various types of extracurricular programs to encourage women's interest in science.  Alongside these innovative programs, increasing awareness of the issues surrounding gender inequity in education have helped to initiate curricular change and draw educators' attention to methods that are most effective in diminishing the gender barrier.  In this section of the paper, I will describe a crossñsection of programs for young women in science.
   The NSF offers grants in two major areas:  Implementation and Development Projects (IDP), and Information Dissemination Activities (IDA).  The IDP goals are primarily to design programs to improve recruitment and retention of women and/or girls in science, engineering, and math careers; to implement these programs based on current research on ways to increase access of women and girls to education and careers; and to encourage permanent changes in the academic, scientific, and social climate for women and girls.  These goals are similar to the goals of many current "intervention programs" for girls and women.  The second area for NSF grants is in IDA.  The goals of these grants is to facilitate dissemination of strategies, research results, and resources which improve the interest and advancement and reduce the barriers of women in science, engineering, and math. (9)  IDA research and work is focused more on helping teachers, administrators, and the public in general understand the problem at hand, where IDP work is the practical result of this academic research, in the form of experiments in new methods or environments.  With the background of IDA type publications, I will now analyze a few IDP type programs.

**GIST Program**

   The first well-documented, concerted effort focused on the problem of girls' disinterest in science was the Girls Into Science and Technology program.  Half of the objective of the GIST program was very similar to, and perhaps a predecessor to, the AAUW Report in 1990.  The GIST project aimed to investigate the underachievement of young women in science in Great Britain, and to combine this research with "action" or "intervention" programs.  It was conducted during the years of 1979-1984, following a cohort of 2000 girls and boys through secondary school.  The primary conclusion from this analytical part of the study was that the problem was not one of girls lacking the opportunity to study science, but one of girls choosing not to study science.  The causes suggested by the GIST report for girls' underachievement in science and technology included "the perceived difficulty of physics, the absence in science studies of social or human implications, girls' relatively lesser experience with scientific and technical toys and games, the expectations for girls' future lives, and the paucity of role models of women in science and engineering". (10)

Popular rhetoric now avoids blaming girls for not choosing science but instead blames institutions, social and educational, that make that choice unlikely. Still, the conclusions drawn by the GIST team of researchers are comparable to some of the AAUW Report recommendations, and are still relevant today as "action" proposals. These chronologically and geographically separated studies reverberate in their call for out-of-school programs featuring female scientist and engineer role models, and handsñon activities to highlight the appeal and relevance of science, perhaps making up for the lack of exposure that girls may have had to science and technology oriented toys.

The most successful of the so-called "intervention programmes" designed by the GIST group to promote their second goal of changing the situation of underachievement of women in the sciences was the VISTA scheme. This scheme involved road shows of working women scientists, students, and technical workers that visited schools to "bring science and technology alive in a way that girls would find sympathetic, and to demonstrate by their presence as role models that women can be feminine, competent, and rewardingly employed in traditionally 'masculine' jobs." (10)

The triumph of the GIST project was that it demonstrated the disparity between potential and their achievement, and raised questions among educators. It does not prescribe specific policies to adopt, but it provides the teachers and schools that would develop those policies with a resource and inspiration to affect change. Since the GIST program and similar developing interest in the United States (especially, again, since the 1992 AAUW study) in improving science education for women, many similar programs have been developed in the United States. Several universities have onñsite summer or weekend programs for elementary, middle, or high school students, since a university is the ideal setting for teaching and labs and enlisting the participation of professors and students as role models. Some programs were sponsored by grants from organizations like the NSF, while others charged the students a fee (usually a bargain because the faculty and student contributions were often voluntary, and many resources of the university could be used).

**WEPAN Organization and WISE Programs**

Many universities have programs for women in science and engineering of various sorts, and the degree of development varies greatly from college to college. Some are groups set up by interested faculty as a part of their career development, others have separate staff dedicated to the specific programs. Most of these programs have many of the same functions, focusing on mentorshipø, colloquia featuring women scientists or about issues women in science face, unique experiences in research or teaching internships, networking and social events, and in many cases, outreach programs to younger girls. In spite of their common goals, these programs are largely independent of one another and there is no "national" Women in Science and

Engineering (WISE) organization. WEPAN, the Women in Engineering Programs and Advocates Network, catalogs many of the programs at universities across the nation. (11) I will select a few of these programs to discuss.

The Women in Engineering Program (WIEP) at Purdue University is the WEPAN Midwest Regional Center, and as such conducts much IDA type work. It also hosts several programs for advocacy of young women in math, science, and engineering. One of the program objectives is to "provide career information and encouragement to preñcollege women to continue interest and achievement in math and science and consider engineering as an appropriate career choice." (11) Its specific programs include Career Days and Future Focus. On Career Days, junior and senior high school women are invited to the Purdue campus with their parents, and are introduced to the universities facilities, faculty, and students. Future Focus programs are like the "road shows" of the GIST program. WIEP members travel to communities throughout the Midwest and offer high school women the opportunity to try hands-on some engineering-related activities, led by female engineering students and practicing engineers. (11) The WIEP programs for high school students are well received, and emphasize some salient characteristics of what works especially well: role modeling, and hands-on activities.

The Stevens Institute of Technology Office of Women's Programs (OWP) is particularly well-organized and has in place several outreach programs. The Exploring Career Options in Engineering and Science Summer Program for grades 10-11 is a comprehensive 2-week summer program featuring hands-on laboratories and research projects, visits to local industry, and panel discussions with professional scientists. A one-day conference, "Women in Engineering and Science: Exploring the World of Discovery", introduces 8-12 grade students, parents, and teachers to technical professions, with hands-on labs and a poster contest celebrating the contributions of women engineers and scientists. Another program involves an OWP staff member and students visiting high schools in the New York and New Jersey area to speak to students about engineering and science disciplines, and their own educational and career paths. One of the student reactions quoted in the description of the program illustrates the impact this might have on a student with less access to professional female role models: "I never met a female engineer until the speaker visited our high school. I always assumed all engineers were male and wore hard hats. She helped me realize that engineering is a great career option for women and that engineers work in a variety of settings." (11)

FLEDGE-ling Camp for Girls at the University of South Florida is a program designed as a collaboration among 3 professors from the departments of Geology, Biology, and Women's Studies. The four-week summer program for 7-8 graders follows the assumption that girls would benefit from a single-sex environment implemented by women scientists, featuring hands-on learning. Accordingly, the program followed the objectives of the National Science Education Standards council and developed a syllabus involving field trips, a group research project, internet research, and Florida ecosystem software programs. This approach addresses effectively many of the current problems in science education for girls: the relative computer illiteracy, the effectiveness of hands-on learning and field trips, the group work as opposed to solely lecture learning, and the presence of female mentors. (12)

The response of the students was very positive. In evaluative surveys, the girls reported a very different view of scientists formed between the first and last day of the program, and even identified themselves as scientists. Their families also reported their daughters' increased confidence in science. (12)

The University of Rochester's own Pre-College Experience in Physics (PREP) is a 4ñweek summer program for 9th and 10th grade high school girls from the Rochester area, in place since 1994. The program is on the University of Rochester campus, and involves interactive lecture and discussion learning, 6 hands-on lab activities, including a written a lab report and oral presentation, 4 major group projects and a day-long activity to promote teamwork and leadership skills among the girls (who arrive not knowing one another), a panel of professionals, visits to several university laboratories including the Laser Lab facility, and numerous speakers and presenters relating their research and work to the topics of study in the program. The program is fast-paced and fun, but rigorous. In presentations and speakers, the focus is on the science, but also the lifestyles of the scientists, and how their careers developed.

The program was very attuned to the positive impact of female role models, and as such was taught by two female undergraduate science students. In evaluative surveys, the students responded positively to the many female role models offered, although some very rationally indicated that they were more concerned with the science than who presented it. (These girls are truly gender-blind!) The projects and laboratories were rated highly in the evaluations, as were the presenters. The interactive lecture style was appreciated, and some students discussed the relative freedom to contribute in the PREP classroom as opposed to coeducational classrooms. The students were a select group of bright and outgoing students, so one might assume that they would not likely to concern themselves with gender issues. However, these students, when asked about school experiences and their feelings on the single-sex program, reacted remarkably maturely in describing the advantages of a single-sex program and were adverse to the idea of having a similar program for both boys and girls (my initial conception of a "fair" program) on the grounds that it would take away from the goals of the PREP Program. These amazingly insightful responses are included as an Appendix. The positive impact that this program had on the girls involved is clear, and their defense of the program's sex exclusivity and their assertion that "other young women should have this experience" indicates an awareness of gender issues at a young age. (14,15) This is probably a result of increasing societal attention to these issues, and is a positive indicator.

These are a few examples of "intervention programs" for women in science. There are numerous others, but many follow along the same general trends as these. The particular programs discussed here were chosen to illustrate the common threads through successful such programs. The AAUW has a database of Math, Science, and Technology Programs for Girls that includes summaries of many high school programs similar to those described above. (16)

As a result of these efforts to level the playing field, the gender gap in the natural sciences in college has narrowed since the AAUW Report in 1992. (17) However, gaps still exist, especially in the area of technology and computing. "We are in the midst of a profound change. As student diversity changes the face of public education, and technology changes the workplace, schools must work smarter and harder to ensure that girls graduate with the knowledge and abilities they need to compete and succeed in the 21$^{st}$-century economy." (18)  In spite of these positive changes, there are still disparities and individual experiences that seem to go back in time.  We are not there yet, in terms of equity in schools.  There is still a place for intervention programs in our society.  The question to address next is, then, what is the most appropriate model for such a program, in light of the progress of the last several years, and engaging the troublesome question of fairness?

**Proposal**

Based on the overwhelmingly positive experience I had organizing and teaching the PREP Program, the concurrence of many of the other out-of-school programs on what aspects of such a program are important and successful, and on the results of a unique survey sent to the PREP participants specifically relevant to this paper, I propose as the best course of action currently a program composed of two major initiatives.  The first part of the program would be a series of science shows at area high schools, and the second part would be a summer day program very similar to the FLEDGE-ling and PREP programs, but broader, including learning a few important parts of many fields (where, for example, the FLEDGE-ling program was focused on geology and the PREP program on physics).  I will call this program Women Into the Natural Sciences (WINS).

There has been progress in narrowing the gender gap in math and sciences in recent years, but my research has indicated that there is still a need for active promotion of the sciences to girls.  As I began to research this subject and think about the ideal program, fairness was a difficult issue for me.  I thoroughly enjoyed the PREP program, but thought the gender exclusivity was unfair to boys.  One mother told me her son would love a program like this too, so "why doesn't the PREP program admit boys also?"  I struggled with this question, thinking that certainly we have progressed enough by now that a mixed sex classroom could be equitable, especially given teachers that are made conscious of the issues, and encouraged to be vigilant.  I considered the fair new solution to be a program similar to PREP but offered to both girls and boys and with conscious focus on diverse role models teaching methods that promote everyone's participation.

Further reading and discussions have muddied the waters for me.  It seems that even today, even the most careful and aware teachers have trouble winning out over the social influences that still cripple girls in the coeducational classroom.  My survey of the girls confirmed this in a way that I never imagined I would see.  These 1998 high school students, children of the generation that declares feminism is dead and tells us that we have destroyed all the barriers, told me that they felt the presence of boys in the classroom would inhibit their learning (15, Appendix).  They commented that the all-female atmosphere had made them more willing to speak out and that the presence of boys would decrease the focus on science.  This is disappointing. It indicates blatantly that we are not there yet, and I therefore propose a single-sex summer program.

The hope continues to be that this is a short-term solution, and that ultimately the resources and staff running this program would shift over to a mixed-sex program simply introducing bright students to science, and to the diversity of careers and people in science.  Simply put, the single-sex program is a fair initiative because as long as society at large and public schools in specific continue to be "affirmative action" for males, a program that is "affirmative action" for females is justified as a counterforce.  As the WINS students will learn, only when there is a balance of external forces can an equilibrium be reached.  Eventually, with progress, the balance will be natural we can work to promote the sciences  in general.

Women Into the Natural Sciences.

**WINS**

I.  Roadshows

Encouraging students at large to get excited about science is a primary focus of many programs nationwide, and this continuing effort is the basis for the first part of this proposed effort.  It would also address the issue of fairness in that it would involve presentations to entire schools, and although conscious consideration would be given in featuring female scientists, no gender exclusivity would be required.

A group of students and/or faculty from the WINS host university would travel to area high schools and present an all-school assembly featuring science demonstrations and informal discussion of what they do in their respective departments.  Ideally, there would also be a few professional scientists on hand to discuss their careers also.  The presenters would not necessarily be exclusively women, but care would be taken to ensure that competent, enthusiastic female scientists were well-represented and were actively "doing" the experiments.  When volunteers from the audience could be used, care would be taken to choose both female and male students.  Some possible experiments:  Luminol (glowing chemical), Van de Graff generator (hair stands up), electrolysis of water and subsequent ignition of $H_2$ gas (loud explosion), optical circuits (break light path in circuit from CD player to speaker and speaker stops), superconductivity (makes things float!).  A script would be generated to be sure that some learning would be incorporated.  Other departments and outreach programs could be consulted for other experiment ideas.  After the assembly, some sort of handout about science activities or local opportunities (science museum, tours) would be passed out (generated by WINS program).  Applications for the summer program would be distributed to these schools, to counselors and teachers, and name recognition from the Roadshows would encourage application.

II.  Summer Program
A.  Schedule and group size

The WINS summer program, either sponsored by a grant or charging a minimal tuition, would be a day program for which students would apply.  A shortcoming of the PREP program was that it could only accept 24 Rochester-area students.  A larger group size is difficult in classrooms and labs.  Therefore I would propose, depending on the facilities available, hiring 4 instructors and accepting a group of 50 that would have separate instruction and lab and project time, and coordinate lessons so that resources could be shared. Presenters could then present to the composite group where appropriate, or in other cases do back-to-back presentations or demos or tours. This would be easier than running the program twice, and would allow the WINS program to reach twice as many young women as most comparable programs. The four week length seemed ideal. To keep students' attention, 9 am to 3 pm would be a good length of day. Some evaluations of the PREP program indicated exhaustion from the 9 am to 4 pm day. (14)

B. Content

Many of the current programs to encourage women in science and engineering have similar basic objectives and methods. The most important aspects of such a program seem to be universally accepted to be:

> 1) Role model female scientists and exposure to a variety of career scientists with different experiences, to show the diversity of science careers, and to open exploration into all of these fields for the students.
>
> 2) Interactive learning, especially hands-on labs and activities with presenters. This will lead to the girls' identifying themselves as scientists or potential scientists.
>
> 3) Group projects/activities to foster collaborative learning and teamwork.
>
> Other elements of a good learning environment that I consider important:
>
> 4) Competent but accessible teachers - ideally undergraduate or graduate students, as opposed to seemingly less "in touch" professors. Confident in their knowledge and ability, but not condescending. First name basis in the classroom and lab.
>
> 5) Focus on teamwork and "icebreaker" activities early on to establish a sense of community and provide a strong basis for later teamwork activities. Just as it is essential to have good collaboration and teamwork in many job situations, it is important that these skills are stressed here.
>
> 6) Frequent and stressed tie-ins of material being learned to everyday phenomena that students may have seen or experienced. As often as possible, demonstrations should be used to illustrate principles.

The program would be structured in Units featuring different areas of science. Within these units, efforts would be made to tie together labs, projects, presenters, and discussion. For example, the unit on Optics might include a presenter showing some interesting new technology, such as a fiber optic circuit and infrared cameras (sense heat, not visible light). The class would have a lecture and discussion briefly introducing optics and the concept of wavelength, and some relevant optics theory. Careers in optics would be discussed, a graduate student from the Optics department would talk about her/his experiences and plans, and the students might do a lab on optical circuit building or lasers (with the relevant theory taught and discussed). Special activities would include visits to local industries or labs, a research project on the Internet on a topic of their choosing, oral and written science presentations and tips on how to do these, a discussion on choosing a college and what colleges are like, and a specific career panel of women to showcase non academic career possibilities in science-related fields. At the conclusion of the program, the students could each prepare a poster project on their favorite topic in science and present it to the combined groups and families at an open house.

**Conclusion**

In order to achieve progress, it is important to understand the historical evolution of an idea or movement. To conceive an appropriate strategy to promote young women into the sciences, we must first assess where we are on the timeline of the "movement" for women in science. We must recognize that changes have begun to occur in the education of women, but we must also concede that there is more work to be done. We define a goal: to have entirely equitable education that does not in any way discourage young women from interest in science, in which case a program exclusively for girls would be as irrelevant and unjustified as a program just for students whose last names begin with K. From the research and current surveys of students, we realize that unfortunately the girls are not yet receiving entirely equitable education (that they are even conscious of this neglect), and we strive to level the playing field somewhat by providing an opportunity for the young women to explore areas historically closed to them in a challenging and eye-opening environment that provides valuable female role models. We implement this program with the expectation that once the young women in our high schools stop reporting indicators of their lack of confidence and encouragement in science, the program will be converted to a program for any bright young student interested in spending a part of their summer learning science and thinking about possible future careers. This program would continue to emphasize not only the diversity of fields and careers in science, but also the diversity of people pursuing these careers.

Anyone agreeing with the practical notion that we need to equip for the increasingly technical world an increasingly well-trained science-abled working force will agree that this program could help us develop more of our "national resources", now and in years to come. With Women Into the Natural Sciences, everybody WINS.

1998.

Appendix 1:  PREP Survey Results, Nov. 1998

**In preliminary questions, all eight respondents said that the PREP program had positively affected the way they approached new in-school or extracurricular science-related activities, and described specific experiences where the physics concepts they had learned, or lab techniques, had helped them in school.**

**All eight also responded very positively to the question:  Do you think other women should have this experience?**

**Question 5: Can you recall experiences in the classroom or lab where a male classmate dominated a project or lab and you or a fellow female classmate were prevented from a hands-on experience?**

Eight respondents said 'no'.

9. Yes.  This year in my biology class, I have a male and a female for lab partners.  After the female and I had gotten the materials for paramecium lab, the male was the first one at the microscope.  Every once in a while he would tell us to look at something.  This same male partner likes to do other labs his way.

**Question 6: In your experience, are male students more aggressive in science courses than females?   "More aggressive" could mean asking more questions demanding more of the teacher's attention, taking over in group projects or labs, etc.**

1&2.  No.

3.  Yes, I have noticed that males tend to ask more questions, but this has not affected me because I tend to ask just as many questions, but I would understand how this obstruction may hinder other females' ability to learn.

4.  I really don't think so.  I have seen both girls and guys doing the stuff described above.  I really don't think the guys do this more, it just depends on how interested the student is in science.

5.  I don't know much about in science courses, but right now I'm in a coed SAT class and I have found that male students are more demanding of the teacher's attention.

6.  Males tend to ask more questions during classtime, but it depends on the individuals in the lab in deciding whoe takes over.  Ex:  In dissecting a rate in 10th grade biology my friend (a girl) and I did all the work even though there were 2 other guys in our group.

7.  In some ways yes and in others no.  I think male students do demand more of a teacher's attention, but I feel that females ask more questions.

8.  No, actually guys seem very immature in school.  I don't think I know of any who are aggressive in science courses.

9.  Generally male students are more aggressive in science courses.  This does vary for each person, though.

**Question 7: Why do you think female students represent much less than half of natural science majors in college when they are more than half of the college population?  This is a difficult question.  Please cite the one or few factors you think contribute most to the underrepresentation of women in science.**

1.  First of all, I was thinking that maybe in the past women were more directed to being secretaries or teachers, but today, it is slowly changing over.  Secondly, maybe women are generally more attracted to the liberal arts for reasons of the way their brain works.

2.  I believe that some women are inclined to feel less important or that they are not as qualified as the men so they don't try for natural science majors.

3.  I find that many females are not interested in science because they don't want to know why things happen.  I find that more males would take apart an engine to learn how it works than a female.  I've also noticed that females tend to gravitate toward humanitarian things, while men tend to do better in math and science.  I've also noticed that females accept facts more, while men ask questions and try to understand why things happen.

4.  Females do not get the same exposure to some fields such as engineering as men do.

5.  I think there is an underrepresentation of women in naturl science majors because they don't have a lot of confidence in the subject.

6.  Science has traditionally been a male-dominated field so people tend to stick to tradition.  Also, education in schools focuses on major figures in science who have accomplished significant things and a majority of these figures are male so it appeals more to males to get involved in scientific fields.

7.  I think some women may shy away from careers in science beause the vast majority of these careers are dominated by males.  It seems like males have always represented much of the science majors so that's the way some people think it should be.  We must change this set stereotype, though.

8.  The female students in college now probably grew up in a world/society that was male dominated, because it was.  But today, as I am growing up, it seems (so far) that girls aand boys are more equal.  (I don't have a job, so I haven't experienced discrimination of being hired at all.)

9.  I think that many women are scared to go into the sciences.  They could be scared of what the men might think of them if they were to have a natural science major.

**Question 8: How do you think the fact that PREP was an all-female environment affected your learning?**

1.  It made me less worried about what others would think about me.  If boys were in it I might feel like acting as if I wanted to be somewhere else in the summer.

2.  I believe that it helped me to feel less intimidated and more open so I learned more.

3.  I don't think it really affected my learning because I would have asked the same questions, and worked the same way if men were there as well, but I found it nice to get to know people who were intellectual, and interested in science, because in school there aren't many people who have the same interests as me.

4.  I did feel more comfortable about some of the stuff we worked on.  I didn't think I would look stupid if I asked a "dumb" question.

6.  It was easier to focus so it was easier to learn, but it didn't affect the degree of my contribution to discussions, labs, etc.

7.  I think the all-female environment helped me learn better.  I felt more confident to try things and say things even if the answer was wrong.

8.  The all-female environment didn't really affect my learning, I don't think.  I think I was able to learn because PREP was fun and interesting, but most importantly, the people surrounding me were friendly as well as interested in learning.

9.  It made me realize that I wasn't the only girl that really likes science and it gave me more confidence.

**Question 9: Was the exposure to women scientists and women mentors in science important?  Why?**

1.  In the way that they could be more of an influence for us, but it didn't affect me that much.

2.  Yes, because it showed us that we are not the only women in science and that we can do just as well as the men.

3.  It was nice to know that there were women scientists in the world, but I found it was not essential to have exposure, I felt that it was not important who was presenting the topic, but rather what the presenting topic was, because I felt the science part was more interesting than the woman part.

4. Yes, I got to see all of the successful women scientists and was inspired.

5. I think the exposure was important because it showed that women could do any profession. They set a good example for young girls.

6. Yes, we don't hear frequently about famous women in science but by hearing about every day women scientists it proved that women can still accomplish things in scientific fields.

7. Yes, I think the exposure to women scientists was important. It helped me to realize that there really are women out there who are scientists (or who had science related careers) and we saw living proof.

8. Somewhat. I think being introduced to different sciences, and finding you enjoy them is the reason you pursue that as a career. But female scientists/mentors did serve as an 'inspiration' to show us that we could do that too!

9. Yes it was, because it shows you that other women have made it, so you can make it too, if you want to.

**Question 12: How do you think the presence of boys would have affected your learning in the PREP program? Assume the program would be identical in structure, presenters, and instructors.**

1. (Referred to question 8)

2. I think I would have not asked as many questions and would have been shyer.

3. I don't think that the presence of boys would have affected my learning, but I feel that a program especially for women is good because the purpose of PREP was to expose women in different science fields; and that if boys were admitted then less females would be given the chance to explore, thus it wouldn't encourage more females into science fields.

4. I don't think that the presence of boys would have affected my learning if the structure would have been identical.

5. I may not have paid enough attention to the lessons if the presence of boys was in the PREP program.

6. I think we wouldn't have been as focused, some boys would have objected to so many female presenters, distracting everyone.

7.  If there were boys involved in PREP I don't think I would have learned the same amount.  The absence of boys helped me to have more confidence and focus more.

8.  The presence of boys might have made some people a little uncomfortable, but if they were interested in having fun and learning then their presence would not have affected my learning.

9.  I probably would not have learned as much if there were boys in the class.  Then it would have been a lot like the situation in school where most boys are more aggressive than girls.

**Question 13: Do you think boys could benefit from being exposed to female AND male career scientists and students, and thinking about issues of balancing career and personal life?**

1.  Yes, because they wouldn't be stereotyping the sciences as being for just men.  Also, maybe they would see that the men don't have to do all the money earning for the family.  Maybe they could help out more in the family.

2.  Yes.

3.  I'm sure they would benefit, because everyone can benefit from a variety of exposures.  Also, by seeing the lifestyles of other scientists, they would be able to conclude what they may want to do in the future.

4.  Yeah, they could.  They would get to see different science fields and see how diversified they were.

5.  I think it would benefit the boys to be exposed to both female and male career scientists and students.  They should also think about the issues of balancing career and personal life.

6.  Yes, they would received same rewards from program, but not to such a great degree.

7.  Yes, they could benefit from that.  I think they should be exposed to scientists of both genders.

8.  Sure.

9.  Yes, I think boys could benefit from it.

**Question 14: React to this proposal:  PREP Program with the same structure, female instructors, and many female role model scientists and students, enrolling 12 female and 12 male students.  Please comment on the "fairness" of such a program, how you think the**

**male students might respond, and how it might affect the females students' experience.**

1. The program would weigh more importance on the women entering into this career area. The male students might respond with the idea that this area of study is more or less for girls. On the other hand, it might influence the girls more. However, it is not "fair" to the boys. Just as today, girls shouldn't get the idea that the natural sciences are more for men.

2. I believe that it would be fair and that the boys would respond but I believe that the girls would be getting less out of the program.

3. This at first may seem fair, but, as I have said, I think that the purpose of PREP was to encourage and expose females to science. However, if half of the class were boys, then the percent of women interested in a science field would go down. I don't think boys should take the spots of females that are unsure of their interest in science. Although this plan may help boys, I feel it would hinder a female further interest in a science career.

4. I don't think this would be very fair to the boys. I think if there were going to be a program like this, it should have male teachers as well. I don't think the female studnets' experience would be affected much except some people may feel uncomfortable and may not be as willing to share their opinions.

5. I think "fairness" might be a small issue. I believe that the female students might have a really good experience. It would probably be better for females than for males. THere might also be the chance of complaints from some of the male students. If it was a really good group of students, however, I think that the class would go well and both male and female students would get a great experience from it!

6. The program is more "fair" because it gives a chance to both males and females, male students would be hesitant to apply (because of female science focus), female students less focused.

7. I think the program is still fair. Seeing that we live in a male dominated world, it wouldn't hurt if females took a little more control (having female instructors, for example) The male students would probably complain about the situation, though. I think the females wouldn't enjoy it as much if males were there.

8. Male students might feel that some male role model/scientists should have been included. I don't think female instructors would bother them, whoever gets the job done best! Having 12 boys and 12 girls seems fair. I don't think it would affect females that much.

9.  The male students might not enjoy all of the people coming in to talk about females in science.  The females and the males might not concentrate as hard on the science aspect of the program either.  Also, it might affect the learning of the females, because the males are more aggressive.